\documentclass[aps,showpacs,pra]{revtex4}
\usepackage{amsmath}
\usepackage{graphicx}
\usepackage{epstopdf}
\usepackage{amsfonts}
\usepackage{amssymb}

\begin{document}
\bibliographystyle{unsrt}

\title{Correlation dynamics for two atoms distributed in two isolated thermal cavities}
\author{Rong-Can Yang$^{1,2}$}
\email{rcyang@fjnu.edu.cn}
\author{Xiu Lin$^{1,2}$}
\author{Heng Ji$^{1,2}$}
\author{Hong-Yu Liu$^{3}$}
\email{liuhongyu@ybu.edu.cn}
\affiliation{$^1$ Fujian Provincial Key Laboratory of Quantum Manipulation and New Energy Materials,\\ College of Physics and Energy, Fujian Normal University, Fuzhou, 350007, China}
\affiliation{$^2$ Fujian Provincial Collaborative Innovation Center for Optoelectronic Semiconductors and Efficient Devices, Xiamen 361005, China}
\affiliation{$^3$ College of Science, Yanbian University, Yanji 133002, China}

\begin{abstract}
In this paper, correlation dynamics for two two-level atoms distributed in two isolated thermal cavities are studied, where the atomic state is initially prepared in a maximum entangled zero-and-two-excitation superposition state. We use two nonclasscal measures including geometric quantum discord via Schatten one norm and concurrence to analyze correlation dynamics for the two atoms when the two cavities have the same or different mean photon numbers. We find though correlation dynamics for these two measures have different features, they have the same characteristics in some particular time when the nonclassical correlation is robust against thermal photon numbers. This result may be important in quantum information processing and quantum computation. 
\end{abstract}


\maketitle

\section{Introduction}
Different from classical physics, superposition exists in quantum, accounting for almost all abnormal behavior in microcosmic fields. When we refer to quantum superposition for two or more than two parties, quantum entanglement, first pointed out by Einstein, Podolsky and Rosen\cite{EPR}, should be referred to. It is a distinct feature of quantum mechanics different from classical one and plays a central role in the test of Bell's theorem without inequalities \cite{Chuang} and viewed as the main feature that gives quantum computers an advantage over their classical counterparts. However in 1998, Knill and Laflamme found that some separable mixed states owning nonclassical correlations but no entanglement can achieve an exponential improvement in efficiency over classical computers for a limited set of tasks\cite{Knill1998}. 

In addition, Ollivier and Zurek \cite{Ollivier2001}, and Henderson and Vedral \cite{Henderson2001} individually found that entanglement is not responsible for all nonclassical correlations and that even separable states usually contain  nonclassical correlations when they analyzed different measures of information in quantum theory. These correlations are aptly named quantum discord. In other words, entanglement is only a strongly nonclassical correlation. Since then, quantum discord has received considerable attention\cite{Laflamme,Datta,Merali}.  However, directly analytical calculation of quantum discord is very hard, even for two-qubit states, since its computing is NP-complete \cite{Huang2014}. This difficulty of extracting analytical solutions led Daki$\acute{c}$, Vedral, and Brukner to propose a geometric measure of quantum discord (GQD) by using the minimal Hilbert-Schmidt norm (or Schatten two-norm) between the given state and a zero-discord state, which only requires a simpler minimization process \cite{Dakic2010}. For the sake of concision, we call it HSGQD. Subsequently, some authors derived lower bound of GQD for any bipartite states and an explicit expression for $2\times n$ systems\cite{Luo2010,Hassan2012,Rana2012,Modi2012}. However, HSGQD has been proved not to be a good measure of quantum correlation because of the increasement under local operations on the unmeasured subsystem \cite{Piani2012}. After that, there are some other GQD measures derived using trace norm (Schatten one-norm, 1NGQD for short)\cite{Debarba2012,Paula2013} and Bures norm \cite{Spehner2013} proposed. Now, it has been shown that 1NGQD is the only p-norm that is able to define a consistent nonclassical correlation measure and the geometry of 1NGQD is equivalent to the quantumness negativity\cite{Nakano2013}. 

Besides, it is all known that distant quantum correlation, especially quantum entanglement is of great importance for distributed quantum computationa, quantum teleportation, and so on\cite{Chuang}. However, Yu and Eberty has showed that two initially entangled qubits without interaction between each other could be later suddenly disentangled completely\cite{TYu2004,TYu2006}. The phenomenon, named entanglement sudden death (ESD) afterwards, is distinctly different from the behavior of local decoherence process, which takes an infinite time evolution under the influence of vacuum fluctuations. Similarly, entanglement can suddenly be generated, which is then called entanglement sudden birth (ESB).

Previously, most of researchers only investigated the hehavior of quantum-correlation (including quantum-entanglement) evolution as the function of time to find some interesting phemenon including ESD and ESB. Different from these proposals, in this paper, we does not only explore quantum-correlation behavior of two distant atoms trapped in two separted thermal cavities, but also study features of the most robust correlation with time gone. We find that the time when the most robust correlation arises is not only a single value but also not proportional to mean photon numbers (MPN) in the two cavities. It emerges as a lader type and reaches a stable time before the phenomenon disappears, i.e. there are no any correlation between the two atoms recreates when MPN is large enough.

The paper is organized as fives sections. Sec.II introduces two measures for nonclassical correlation (NCC) including 1NGQD and concurrence. We then give our model and numerious simulation in the third and fourth sections, respectively. Conclusion is made in Sec.V. 

\section{nonclassical measures for a bipartite system}
In this section, let us introduce two different measures to quantify nonclassical correlation for bipartite systems formed by two 2-dimensional subsystems. One is 1NGQD to calculate quantum correlation, and the other one is Wootters concurrence (also called concurrence for short) quantifying the degree of quantum entanglement for any two-qubit state.
\subsection{1NGQD}
According to Ref.\cite{Debarba2012,Paula2013}, for a bipartite system (A and B) with density matrix $\rho_{AB}$, its 1NGQD is defined as the minimal (trace norm) distance between $\rho_{AB}$ and its closest classical state $\chi_{\rho}$, i.e. 
\begin{equation}
D_1(\rho_{AB})=\min_{\chi_{\rho} \in \Omega_0}(\left|\left|\rho_{AB}-\chi_{\rho}\right|\right|_1)
\end{equation}
with $\Omega_0$ is the set of classical quantum states which have zero quantum discord with respect to local measurements on one subsystem A and $\left|\left|M\right|\right|_1 = tr(\sqrt{M^+M})$ is the Schatten trace norm (one norm). When the density matrx $\rho_{AB}$ is written in its Bloch representation as 
\begin{equation}\label{Bloch}
\rho_{AB}=\frac{1}{4}\left[I^A\otimes I^B + \sum_{j=1,2,3} (x_j\sigma_j \otimes I^B + y_j I^A \otimes \sigma_j) +\sum_{j,k=1,2,3}t_{jk}\sigma_j \otimes\sigma_k\right]
\end{equation}
where $x_j=tr\left(\rho_{AB}(\sigma_j \otimes I^B)\right)$ is the element of the local Bloch vector $\vec{X}$ and $\{\sigma_j\}$ are the Pauli spin matrices.  If $\rho_{AB}$ has the X-class form, i.e.
\begin{equation}\label{Xclass}
\rho_{AB}=\left(
\begin{matrix}
a     & 0    & 0 & w\\
0     & b    & z & 0\\
0     & z^*& c & 0\\
w^*& 0    & 0 & d
\end{matrix}
\right)
\end{equation}
with $z^* (w^*)$ being the complex conjugate of $z (w)$, then we can easily obtain \cite{Hu2018,Mohamed2018,Obando2018}
\begin{equation}\label{XYT}
\begin{split}
\vec{X}&=\left(0,0,a+b-c-d\right)^t,\\
\vec{Y}&=\left(0,0,a-b+c-d\right)^t,\\
T=(t_{ij})=&\left(
\begin{matrix}
2\text{Re}(w+z)  &  2\text{Im}(z-w)  &  0\\
-2\text{Im}(w+z)  &  2\text{Re}(z-w)  &  0\\
0                        &  0                       &  a-b-c+d
\end{matrix}
\right).\\
\end{split}
\end{equation}
According to Ref.\cite{Ciccarello2014,Mohamed2018}, we can calculate the 1NGQD for the bipartite system with the formular
\begin{equation}
D_1(\rho_{AB})=\frac{1}{2}\sqrt{\frac{t_{11}^2f_1-t_{22}^2f_2}{f_1-f_2+t_{11}^2-t_{22}^2}}
\end{equation}
with $f_1=\max\left(t_{33}^2,t_{22}^2+x_3^2\right), f_2=\min\left(t_{33}^2,t_{11}^2\right)$. In order to compare its value with the concurrence, we \textbf{double 1NGQD} in the following passages, i.e. $D_1(\rho_{AB})=2D_1(\rho_{AB})$.

\subsection{Concurrence}
Concurrence is a well-konwn entanglement measure for bipartite systems \cite{Wootters1998}. For a  bipartite system with density matrix $\rho_{AB}$, its concurrence is defined as
\begin{equation}\label{concurrence}
C(\rho_{AB})=\max(0,\lambda_1-\lambda_2-\lambda_3-\lambda_4)
\end{equation}
with $\lambda_j (j=1,2,3,4)$ being the square root of the eigenvalues of the non-Hermitian matrix $\rho_{AB}(\sigma^y\otimes \sigma^y)\rho_{AB}^*(\sigma^y\otimes \sigma^y)$ in decreasing order and $\rho_{AB}^*$ the complex conjugation of $\rho_{AB}$ in the standard basis. If the density matrix $\rho_{AB}$ has X-class form described in Eq.(\ref{Xclass}), then its concurrence can be simplified as \cite{TYu2007}
\begin{equation}
C(\rho)=2\max(0, \left|z\right|-\sqrt{ad},\left|w\right|-\sqrt{bc})
\end{equation}

\section{Model}

\begin{figure}
\centering
  \includegraphics[width=.6\textwidth]{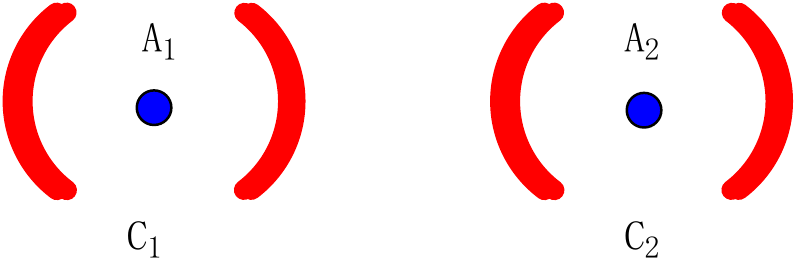}
  \caption{Sketch for two atoms distributed in two isolated thermal cavities}
  \label{model}
\end{figure}

We suppose that there are two identical two-level atoms (marked by $A_1$ and $A_2$) individually trapped in a single-mode cavity  ($C_1$ and $C_2$), shown in Fig.\ref{model}. With consideration of resonant interaction between atoms and cavities, interaction Hamiltonian for the systems can be written as ($\hbar=1$)
\begin{equation}\label{Hamiltonian}
H=\sum\limits_{j=1}^2 (g_j \sigma_j^+ a_j  +H.c.),
\end{equation}
where the subscript $j$ denotes the $\mathit{j}$-th cavity (atom), $g_j$ represents the atom-cavity coupling rate; $\sigma^+=\left|e\right>\left<g\right|$ with $\left|e(g)\right>$ being atomic excited (ground) state and $a$ are separetely lowering operator for atoms and annihilation operator for cavities and $H.c.$ Hermite terms.

Consider both the cavities $C_1$  and $C_2$ are initially prepared in the thermal state with MPN $\bar{n}_1$ and $\bar{n}_2$, respectivity, i.e.
\begin{equation}\label{themalstate}
\rho_{C_{1(2)}}=\sum\limits_{n=0}^{\infty} \frac{\bar{n}_{1(2)}^n}{(1+\bar{n}_{1(2)})^{n+1}}\left|n\right>_{1(2)}\left<n\right|,
\end{equation}
while the two atoms are initially in a maximum entangled state $\left|\phi_{A_1A_2}\right>=\frac{1}{\sqrt{2}}(\left|ee\right>_{12}+\left|gg\right>_{12})$. Thus, the initial density operator of the system is $\rho_S(0)=\rho_{C_1}\otimes\rho_{C_2}\otimes \left|\phi_{A_1A_2}\right> \left<\phi_{A_1A_2}\right|$. Under the action of the Hamiltonian (\ref{Hamiltonian}), the system state at any time $t$ will evolve to 
\begin{equation}
\begin{split}
\rho_S(t)&=e^{-iHt}\rho(0)e^{iHt}\\
&=\frac{1}{2}\sum\limits_{m,n=0}^{\infty} \frac{\bar{n}_1^m}{(1+\bar{n}_1)^{m+1}} \frac{\bar{n}_2^n}{(1+\bar{n}_2)^{n+1}}\left[\cos(\sqrt{m+1}g_1t)\cos(\sqrt{n+1}g_2t)\left|ee\right>\left|m,n\right>\right.\\
&-i \cos(\sqrt{m+1}g_1t)\sin(\sqrt{n+1}g_2t)\left|eg\right>\left|m,n+1\right>-i \sin(\sqrt{m+1}g_1t)\cos(\sqrt{n+1}g_2t)\left|ge\right>\left|m+1,n\right>\\
&+\sin(\sqrt{m+1}g_1t)\sin(\sqrt{n+1}g_2t)\left|gg\right>\left|m+1,n+1\right>+\cos(\sqrt{m}g_1t)\cos(\sqrt{n}g_2t)\left|gg\right>\left|m,n\right>\\
&-i \cos(\sqrt{m}g_1t)\sin(\sqrt{n}g_2t)\left|ge\right>\left|m,n-1\right>-i \sin(\sqrt{m}g_1t)\cos(\sqrt{n}g_2t)\left|eg\right>\left|m-1,n\right>\\
&\left.+\sin(\sqrt{m}g_1t)\sin(\sqrt{n}g_2t)\left|ee\right>\left|m-1,n-1\right>\right]\\
&\times \left[\cos(\sqrt{m+1}g_1t)\cos(\sqrt{n+1}g_2t)\left<ee\right|\left<m,n\right| +\sin(\sqrt{m+1}g_1t)\sin(\sqrt{n+1}g_2t)\left<gg\right|\left<m+1,n+1\right|\right.\\
&+i \cos(\sqrt{m+1}g_1t)\sin(\sqrt{n+1}g_2t)\left<eg\right|\left<m,n+1\right|+i \sin(\sqrt{m+1}g_1t)\cos(\sqrt{n+1}g_2t)\left<ge\right|\left<m+1,n\right|\\
&+\cos(\sqrt{m}g_1t)\cos(\sqrt{n}g_2t)\left<gg\right|\left<m,n\right| +\sin(\sqrt{m}g_1t)\sin(\sqrt{n}g_2t)\left<ee\right|\left<m-1,n-1\right|\\
&\left.+i \cos(\sqrt{m}g_1t)\sin(\sqrt{n}g_2t)\left<ge\right|\left<m,n-1\right|+i \sin(\sqrt{m}g_1t)\cos(\sqrt{n}g_2t)\left<eg\right|\left<m-1,n\right|  \right]\\
\end{split}
\end{equation}
After partially tracing the two-cavity state, we find that atomic density matrix at any time $t$ ($\rho_{A_1A_2}(t)=tr_{C_1C_2}\rho_S(t)$)  has a X-class form described in Eq. (\ref{Xclass}) with
\begin{equation}
\begin{split}
&a=\frac{1}{2}\sum\limits_{m,n=0}^{\infty} \frac{\bar{n}_1^m}{(1+\bar{n}_1)^{m+1}} \frac{\bar{n}_2^n}{(1+\bar{n}_2)^{n+1}}\left(\cos^2\sqrt{m+1}g_1t\cos^2\sqrt{n+1}g_2t + \sin^2\sqrt{m}g_1t\sin^2\sqrt{n}g_2t\right),\\
&b=\frac{1}{2}\sum\limits_{m,n=0}^{\infty} \frac{\bar{n}_1^m}{(1+\bar{n}_1)^{m+1}} \frac{\bar{n}_2^n}{(1+\bar{n}_2)^{n+1}}\left(\cos^2\sqrt{m+1}g_1t\sin^2\sqrt{n+1}g_2t + \sin^2\sqrt{m}g_1t\cos^2\sqrt{n}g_2t\right),\\
&c=\frac{1}{2}\sum\limits_{m,n=0}^{\infty} \frac{\bar{n}_1^m}{(1+\bar{n}_1)^{m+1}} \frac{\bar{n}_2^n}{(1+\bar{n}_2)^{n+1}}\left(\sin^2\sqrt{m+1}g_1t\cos^2\sqrt{n+1}g_2t + \cos^2\sqrt{m}g_1t\sin^2\sqrt{n}g_2t\right),\\
&d=\frac{1}{2}\sum\limits_{m,n=0}^{\infty} \frac{\bar{n}_1^m}{(1+\bar{n}_1)^{m+1}} \frac{\bar{n}_2^n}{(1+\bar{n}_2)^{n+1}}\left(\sin^2\sqrt{m+1}g_1t\sin^2\sqrt{n+1}g_2t + \cos^2\sqrt{m}g_1t\cos^2\sqrt{n}g_2t\right),\\
&w=\frac{1}{2}\sum\limits_{m,n=0}^{\infty} \frac{\bar{n}_1^m}{(1+\bar{n}_1)^{m+1}} \frac{\bar{n}_2^n}{(1+\bar{n}_2)^{n+1}}\cos\sqrt{m+1}g_1t\cos\sqrt{n+1}g_2t\cos\sqrt{m}g_1t\cos\sqrt{n}g_2t,\\
&z=0.
\end{split}
\end{equation}
If we express $\rho_{A_1A_2}(t)$ in its Bloch representation described in Eq.(\ref{Bloch}), then we can obtain its local vectors $\vec{X},\vec{Y}$ and correlation matrix $T$, i.e.
\begin{equation}
\begin{split}
&\vec{X}=2\left(0,0,a+b-1/2\right)^t,\\
&\vec{Y}=2\left(0,0,a+c-1/2\right)^t,\\
&T=(t_{jk})=2\left(
\begin{matrix}
w & 0 & 0 \\
0 & -w & 0 \\
0 & 0 & a+d-\frac{1}{2}
\end{matrix}
\right).\\
\end{split}
\end{equation}

According to the definition of 1NGQD and concurrence described in Sec. II, we can easily obtain their forms for our model, i.e.
\begin{equation}
\begin{split}
Concurrence&: \qquad C(\rho_{A_1A_2})=2\max(0,\left| w \right| - \sqrt{bc}),\\
1NGQD&: \qquad D_1(\rho_{A_1A_2})=\left|w\right| \\
\end{split}
\end{equation}

\section{numerical simulation}
\begin{figure}
\centering
  \includegraphics[width=.6\textwidth]{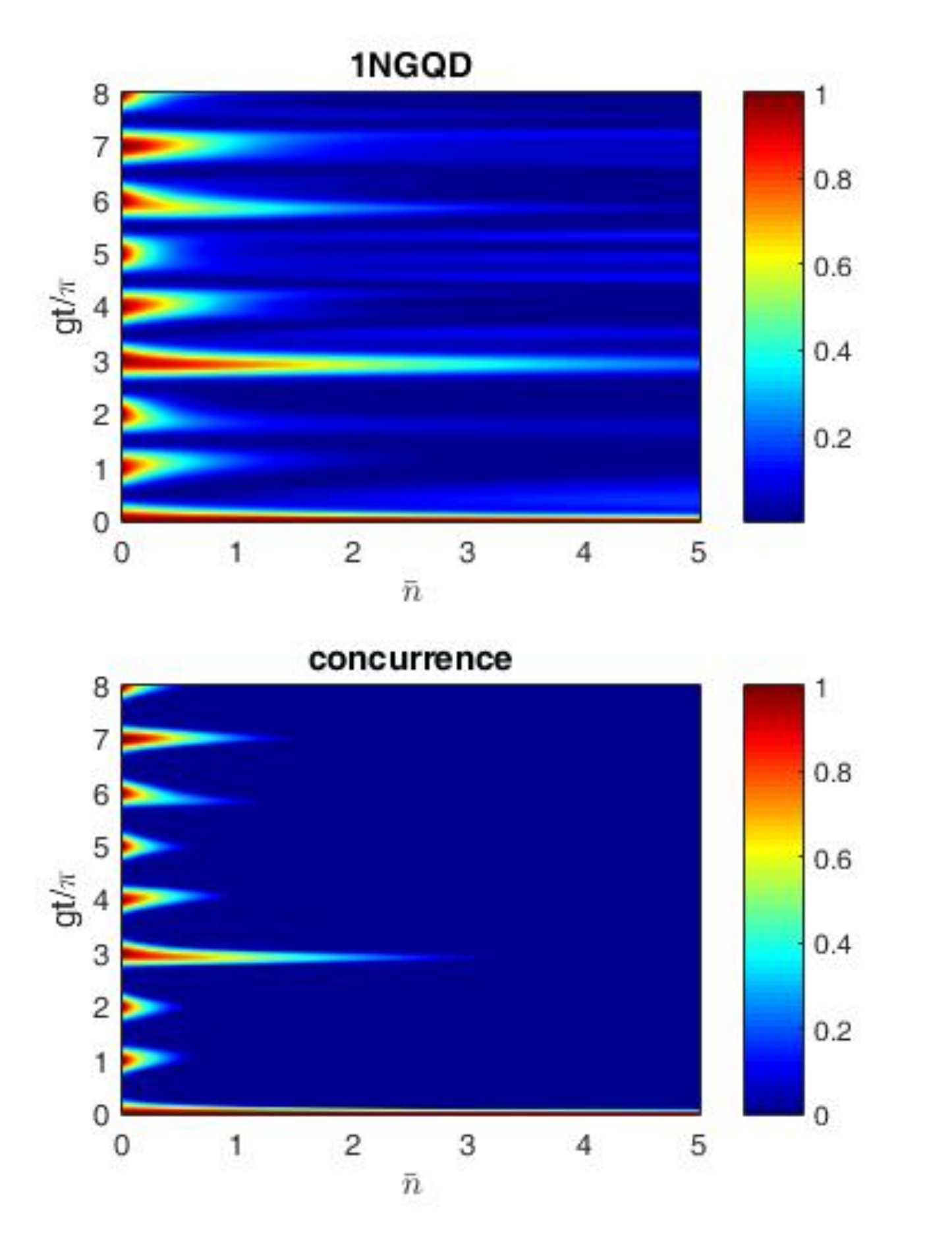}
  \caption{1NGQD (top) and Concurrence (bottom) as the function of non-dimensional time $gt$ and MPN $\bar{n}_1=\bar{n}_2=\bar{n}$ with $g_1=g_2=g$.}
  \label{correlation_3D_n1=n2_20181126_GQD_and_Con_3D}
\end{figure}

In this part, we will use numberical simulation to study 1NGQD and concurrence. For the sake of simplicity, we set $g_1=g_2=g$ in the following passages. Since correlations are influenced by both MPNs in two cavities, we first consider $\bar{n}_1=\bar{n}_2=\bar{n}$. 1NGQD and concurrence for the two atoms as the function of non-dimensional time $gt$ and $\bar{n}$ are illustrated in Fig.\ref{correlation_3D_n1=n2_20181126_GQD_and_Con_3D}. It is clearly shown that both measuring results are distinctly different from each other except when both two cavities are initially in vacuum states ($\bar{n}=0$). When $\bar{n}=0$, $D_1(\rho_{A_1A_2})=\left|w\right|=\cos^2\left(gt\right)$ and  $C(\rho_{A_1A_2})=\cos^4\left(gt\right)$ can be easily obtained. Thus, they are both periodically time-dependent with periodicity $\pi$ and at the same time decrease from maximum value 1 to mimum value zero, and then increase to one again. However, when $\bar{n} \neq 0$, 1NGQD and concurrence show distinctly different features. We can find that concurrence quickly descreases to zero and remains zero for some while and then rebirthes, while 1NGQD is only influenced at its magnitudes  when $\bar{n}$ is small. In other words, ESD and ESB exist for entanglement \cite{TYu2007}, while 1NGQD has no this characteristic. Of course, all correlations including 1NGQD and concurrence gradually disappear as the increase of $\bar{n}$. Besides, the peak of 1NGQD and concurrence in each period arises near $gt=k\pi (k=1,2,3,...)$ when $\bar{n}$ (including $\bar{n}=0$) is small. Especially, there is a special non-dimentional time $g\tau$ near $3\pi$ when correlation including 1NGQD and concurrence is much larger than that of the other rebirth point. Similarly, we can obtain similar features in the case of $\bar{n}_1 \neq \bar{n}_2$.

\begin{figure}
	\centering
	\includegraphics[width=.6\textwidth]{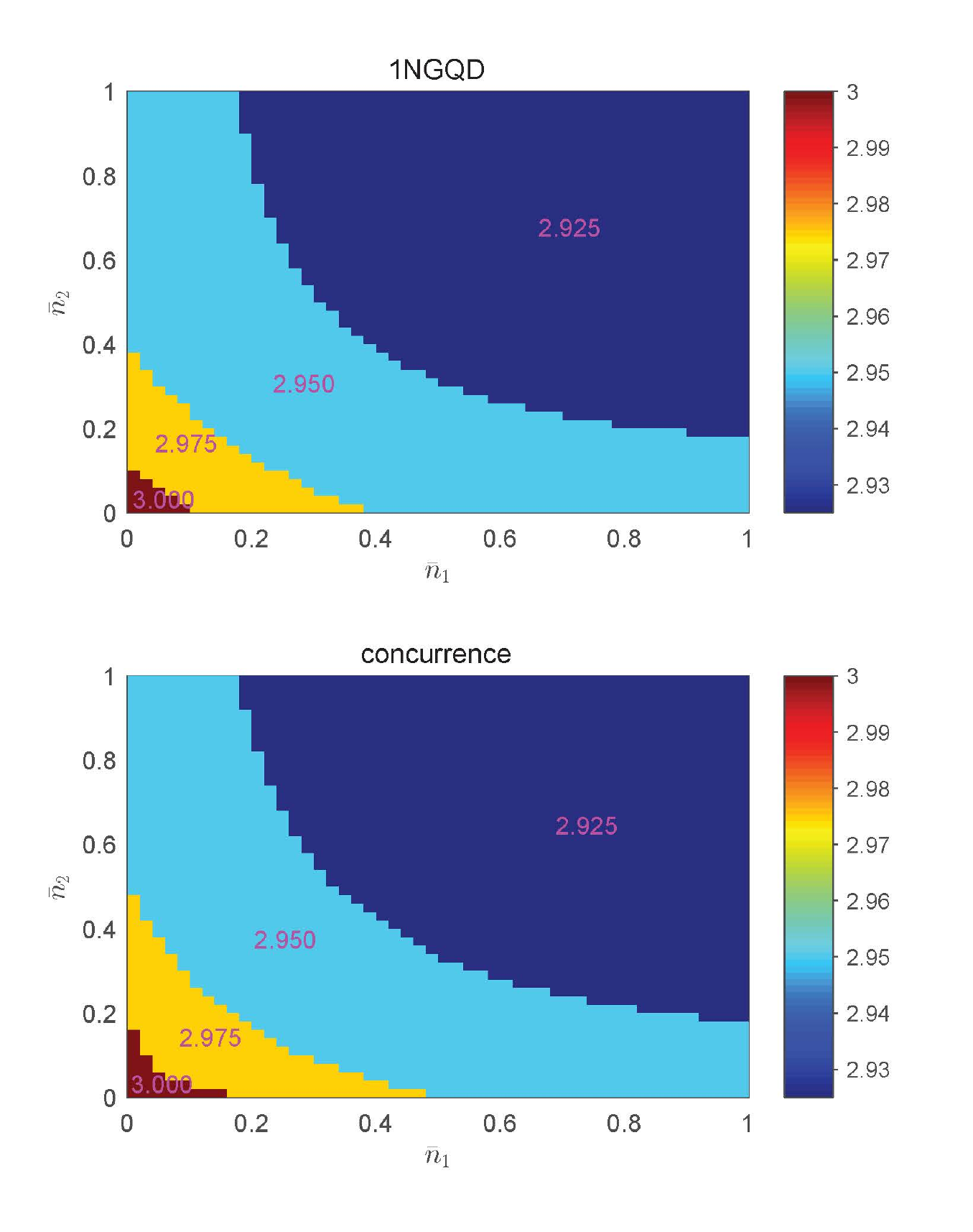}
	\caption{Non-dimensional time $g\tau/\pi$ for 1NGQD (top) and concurrence (bottom) as the function of $\bar{n}_1$ and $\bar{n}_2$ when a robust correlation arises with $g_1=g_2=g$.}
	\label{correlation_20181126_GQD_and_Con_meanphoton_3D}
\end{figure}

In order to determin $g\tau$, we give the illustration for the relation between the non-dimensional time $g\tau/\pi$ as the function of $\bar{n}_1$ and $\bar{n}_2$, shown in Fig.\ref{correlation_20181126_GQD_and_Con_meanphoton_3D}. Based on these two figures, we can clearly see that $g\tau/\pi$ can only be chosen within four decrete values $3.000, 2.975, 2.950$ and $2.925$ for 1NGQD and concurrence no matter $\bar{n}_1 = \bar{n}_2$ or $\bar{n}_1 \neq \bar{n}_2$. For 1NGQD (concurrence), both $\bar{n}_1$ and $\bar{n}_2$ are less than about $0.1 (0.15)$ and $0.4 (0.5)$, $g\tau/\pi$ can be chosen as $3.000$ and $2.975$, respectively. While the boudary for $g\tau/\pi=2.950$ and $2.925$, both 1NGQD and concurrence are nearly the same.

\section{Conclusion}
In conclusion, correlation dynamics for two atoms distributed in two separated thermal cavities is studied,where two measures including 1NGQD and concurrence are used. It is not only proved that non-classical correlation exists even if there is no entanglement, but also shows that ESD and ESB happen only when neither the two thermal cavities are in vacuum state. In addition, this phenomenon is more obvious with the increase of the mean photon number when it is not very large. While there are no any sudden death for 1NGQD, which is only affected at its magnitudes. Most of importance, an interesting phenomenon is found, where a robust correlation can only be obtained at descrete time satisfying $g\tau/\pi=3.000, 2.975, 2.950$ and $2.925$ even when MPNs in the two cavities are not both zero or very large. We hope that it may be important for quantum information processing and quantum computation.
\section{Acknowledgments}
   This work is supported by the National Natural Science Foundation of Fujian (Grant no. 2018J01661) and the outstanding Young Talent Fund Project of Jilin Province (Grant No.20180520223JH).


\end{document}